\documentclass[aps,prd,twocolumn,superscriptaddress,showkeys]{revtex4-1}  

\usepackage{graphicx}
\usepackage{subfigure}
\usepackage{epstopdf}
\usepackage{rotating}
\usepackage{scalerel}
\usepackage[colorlinks,linkcolor=blue,anchorcolor=blue,citecolor=blue]{hyperref}

\usepackage{comment}

\usepackage{amsmath}
\usepackage{amsfonts}
\usepackage{amsthm}
\usepackage{amsbsy}
\usepackage{amssymb}
\usepackage{bm}
\usepackage{braket}
\usepackage{cancel}
\usepackage{mathtools}
\usepackage{upgreek}
\usepackage{wasysym}
\usepackage[version-1-compatibility]{siunitx}
\newcommand{\electron}{e^-}
\newcommand{\positron}{e^+}
\newcommand{\epem}{\positron\electron}

\newcommand{\kp}{K^+}
\newcommand{\km}{K^-}
\newcommand{\ks}{K_S^0}
\newcommand{\kaon}{K}
\newcommand{\kaonb}{\bar{K}}

\newcommand{\pip}{\pi^+}
\newcommand{\pim}{\pi^-}

\newcommand{\kpkm}{\kp\km}
\newcommand{\ksks}{\ks\ks}
\newcommand{\kkbar}{\kaon\kaonb}

\newcommand{\phikkbar}{\phi\kaon\kaonb}
\newcommand{\kpkmkpkm}{\kpkm\kpkm}
\newcommand{\kpkmksks}{\kpkm\ksks}

\newcommand{\epemtophikpkm}{\epem \to \phi \kpkm}
\newcommand{\epemtophiksks}{\epem \to \phi \ksks}

\newcommand{\proton}{p}
\newcommand{\antiproton}{\bar{p}}
\newcommand{\ppbar}{\proton\antiproton}

\usepackage{dcolumn}
\usepackage{color}
\usepackage{overpic}
\usepackage{xspace}
\usepackage{textpos}
\usepackage[english]{babel}
\usepackage{accents}

\usepackage{lineno}

\newcommand{\jpsi}{J/\psi}
\newcommand{\gevcc}{\mathrm{GeV}/c^2}

\newcommand{\gev}{\mathrm{GeV}}
\newcommand{\mev}{\mathrm{MeV}}

\newcommand{\chisq}{\chi^2}
\newcommand{\X}{\chi_{c1}(3872)}

\newcommand{\Yl}{\psi(4230)}
\newcommand{\Yh}{\psi(4360)}
\newcommand{\Zc}{Z_c(3900)}
\newcommand{\isr}{\gamma_{\strut\mathrm{ISR}}}
\newcommand{\pipi}{\pi^+\pi^-}
\newcommand{\ee}{e^+e^-}

\newcommand{\etal}{{\it et al. }}

\begin{document}

\linenumbers


\title{Cross section measurements of $\boldmath{\epemtophikpkm}$ and $\boldmath{\epemtophiksks}$ at center-of-mass energies between 3.7730 GeV and 4.7008 GeV}

\author{
  M.~Ablikim$^{1}$, M.~N.~Achasov$^{5,b}$, P.~Adlarson$^{75}$, X.~C.~Ai$^{81}$, R.~Aliberti$^{36}$, A.~Amoroso$^{74A,74C}$, M.~R.~An$^{40}$, Q.~An$^{71,58}$, Y.~Bai$^{57}$, O.~Bakina$^{37}$
% ...
\\
\vspace{0.2cm}
(BESIII Collaboration)\\
\vspace{0.2cm} {\it
	$^{1}$ Institute of High Energy Physics, Beijing 100049, People's Republic of China\\
% ...
$^{5}$ Budker Institute of Nuclear Physics SB RAS (BINP), Novosibirsk 630090, Russia\\
% ...
$^{36}$ Johannes Gutenberg University of Mainz, Johann-Joachim-Becher-Weg 45, D-55099 Mainz, Germany\\
	$^{37}$ Joint Institute for Nuclear Research, 141980 Dubna, Moscow region, Russia\\
% ...
$^{40}$ Liaoning Normal University, Dalian 116029, People's Republic of China\\
% ...
$^{57}$ Southeast University, Nanjing 211100, People's Republic of China\\
	$^{58}$ State Key Laboratory of Particle Detection and Electronics, Beijing 100049, Hefei 230026, People's Republic of China\\
% ...
$^{71}$ University of Science and Technology of China, Hefei 230026, People's Republic of China\\
% ...
$^{74}$ University of Turin and INFN, (A)University of Turin, I-10125, Turin, Italy; (B)University of Eastern Piedmont, I-15121, Alessandria, Italy; (C)INFN, I-10125, Turin, Italy\\
	$^{75}$ Uppsala University, Box 516, SE-75120 Uppsala, Sweden\\
% ...
$^{81}$ Zhengzhou University, Zhengzhou 450001, People's Republic of China\\
	\vspace{0.2cm}
% ...
$^{b}$ Also at the Novosibirsk State University, Novosibirsk, 630090, Russia\\
% ...
}
}

\date{\today}

\begin{abstract}
  Based on \SI{22.7}{fb^{-1}} of $\epem$ annihilation data collected at 33 different center-of-mass energies between \SI{3.7730}{GeV} and \SI{4.7008}{GeV} with the BESIII detector at the BEPCII collider, Born cross sections of the two processes $\epemtophikpkm$ and $\epemtophiksks$ are measured for the first time.
No indication of resonant production through an intermediate vector state $V$ is observed, and 
the upper limits on the product of the electronic width $\Gamma_{\epem}$ and the branching fraction $Br(V\rightarrow \phi\kkbar)$ of the processes $\epem \to V \to \phi \kpkm$ and $\epem \to V \to \phi \ksks$ at the \SI{90}{\%} confidence level are obtained for a large parameter space in resonance masses and widths.
For the current world average mass and width of the $\psi(4230)$ of $m=4.2187~\gevcc$ and $\Gamma=44~\mev$, we set upper limits on the $\phi\kpkm$ and $\phi\ksks$ final states of $1.75~\textrm{eV}$ and $0.47~\textrm{eV}$ at the $90\%$ confidence level, respectively.
\end{abstract}


\maketitle

\section{\label{sec:intro} Introduction}

In the past several years, many exotic candidates have been discovered in the charmonium and charmonium-like spectrum. Notable examples are the $\X$ discovered by Belle \cite{Choi:2003ue}, the charged charmonium-like $\Zc$ discovered by BESIII \cite{Ablikim:2013mio}, and the $Y(4260)$ originally observed by BaBar \cite{Aubert:2005rm} as a single broad peak in the $\ee\to\isr \pipi \jpsi$ process, where ISR denotes initial state radiation. 
BESIII later revealed that the broad $Y(4260)$ peak is asymmetric and fit it with two resonances, one (the $\Yl$) with a slightly lower mass than the $Y(4260)$ and one (the $\Yh$) with a higher mass~\cite{ref1}.

The $\Yl$ has been clearly observed by BESIII in prominent charmonium transitions to $\pipi\jpsi$~\cite{ref1}, $\pipi h_c$~\cite{ref2}, $\pipi \psi(3686)$~\cite{ref3}, $\omega\chi_{c0}$~\cite{ref5} and $\eta\jpsi$~\cite{bes3jpsieta1, bes3jpsieta2, ref6}. However, decays into light hadrons have not been so far observed, including $\ppbar \pi^0$~\cite{ref7}, $\phi\phi\phi$, $\phi\phi\omega$~\cite{ref8}, $p K_S^0 \bar{n} K^-$~\cite{ref9}, $K_S^0 K^\pm \pi^\mp$~\cite{ref10}, $K_S^0 K^\pm \pi^\mp \pi^0$, $K_S^0 K^\pm \pi^\mp \eta$~\cite{ref11}, $2(p\bar{p})$~\cite{ref13}, $\phi\Lambda\bar{\Lambda}$~\cite{ref14}, $p\bar{p}\eta$ and $p\bar{p}\omega$~\cite{ref15}.

While multiple theoretical approaches have attempted to classify exotic states, e.g. as tetraquarks, hadronic molecules or hybrid charmonia, there is still no full understanding of the inner structure of, e.g., the $\Yl$. Their compatibility with experimental data has recently been discussed in detail in Ref.~\cite{brambilla}. Additional sources of information, including new decay modes of the $\Yl$, are needed from experiments in order to discriminate between the different hypotheses. In Ref.~\cite{ref16}, the $\Yl$ is interpreted as a diquark antidiquark state $cs\bar{c}\bar{s}$, which would lead to decays into final states containing $s\bar{s}$. In fact, this was supported by \mbox{BESIII} in a spin-parity analysis of the $\Zc$ in the process $\epem\to \Yl \to \Zc \pi \to \pipi\jpsi$~\cite{y_to_ssbar}, with one of the dominant contributions coming from $\epem\to\Yl\to f_0(980)\jpsi$. The $f_0(980)$ meson is known to have large $s\bar{s}$ contributions~\cite{f980_ssbar1,f980_ssbar2}. Assuming that the $c\bar{c}$ component of the $cs\bar{c}\bar{s}$ state annihilates while the $s\bar{s}$ survives as a meson with hidden strangeness, e.g. the $\phi$, the decay $\epem\to\Yl\to\phi\kaon\kaonb$ is expected to occur. Further analyses indicating $s\bar{s}$ components of the $\Yl$ can be found in~\cite{Y_has_ssbar1,Y_has_ssbar2,Y_has_ssbar3}.

In this work, the measurements of the energy-dependent Born cross sections of the processes $\epemtophikpkm$ and $\epemtophiksks$ for data collected at 33 different center-of-mass energies between $3.7730~\gev$ and 4.7008~$\gev$ with the BESIII detector are reported. Possible resonant contributions $V\to\phi \kpkm$ and $V\to\phi \ksks$ are investigated.

\section{\label{sec:bes} BESIII Detector and Monte Carlo Simulations}

The BESIII detector is a magnetic spectrometer~\cite{Ablikim:2009aa} located at the Beijing Electron Positron Collider (BEPCII)~\cite{Yu:IPAC2016-TUYA01}.
The cylindrical core of the \mbox{BESIII} detector consists of a helium-based multilayer drift chamber (MDC), a plastic scintillator time-of-flight system (TOF), and a CsI(Tl) electromagnetic calorimeter (EMC), which are all enclosed in a superconducting solenoidal magnet providing a 1.0~T magnetic field~\cite{detector_nachtrag}.
The solenoid is supported by an octagonal flux-return yoke with resistive plate counter muon identifier modules interleaved with steel.
The acceptance for charged particles and photons is 93\% over the $4\pi$ solid angle.
The charged-particle momentum resolution at $1~{\rm GeV}/c$ is $0.5\%$, and the $dE/dx$ resolution is $6\%$ for the electrons from Bhabha scattering.
The EMC measures photon energies with a resolution of $2.5\%$ ($5\%$) at $1$~GeV in the barrel (end cap) region.
The time resolution of the TOF barrel part is 68~ps, while that of the end cap part is 110~ps.
The end cap TOF system was upgraded in 2015 with multi-gap resistive plate chamber technology, providing a time resolution of 60~ps~\cite{etof}.  This upgrade improves the data taken at 27 of the 33 center-of-mass energy points.  

A Monte Carlo (MC) simulation of the BESIII detector including a realistic representation of the electronic readout, based on \textsc{geant}{\footnotesize 4}~\cite{geant4}, is used to optimize particle selection requirements, to determine the product of detector acceptance and reconstruction efficiency, and to study and estimate possible background contributions. These simulations also account for the observed beam energy spread.

Dedicated simulations with $2.5 \times 10^5$ events per center-of-mass energy of the signal processes $\epemtophikpkm$ and $\epemtophiksks$ with subsequent decays $\phi\to\kpkm$ and $\ks\to\pip\pim$ are generated with the {\sc KKMC}~\cite{ref:kkmc} generator, accounting for ISR and vacuum polarization (VP). 

In addition, an inclusive MC sample simulated for a center-of-mass energy of $4.1784~\gev$, corresponding to the dataset with the largest integrated luminosity (see Table~\ref{tab:xsec}), is used to study potential background contributions. This sample includes open charm processes, ISR production of vector charmonium(-like) states and continuum $q\bar{q}$ (where $q$ is a $u,d,s$ quark) processes. Known decay modes are modeled with {\sc evtgen}~\cite{ref:evtgen} using branching fractions taken from the Particle Data Group (PDG)~\cite{pdg}, whereas unknown processes are modeled by the {\sc lundcharm} model~\cite{ref:lundcharm}. Final state radiation from charged final state particles is incorporated with the {\sc photos} package~\cite{photos}.
The inclusive MC sample at $\sqrt{s}=4.1784~\gev$ corresponds to 40 times the luminosity for the data taken at this center-of-mass energy.

\section{\label{sec:sel}Event selection}
\begin{figure}[tbh!]
\begin{overpic}[width=0.4\textwidth]{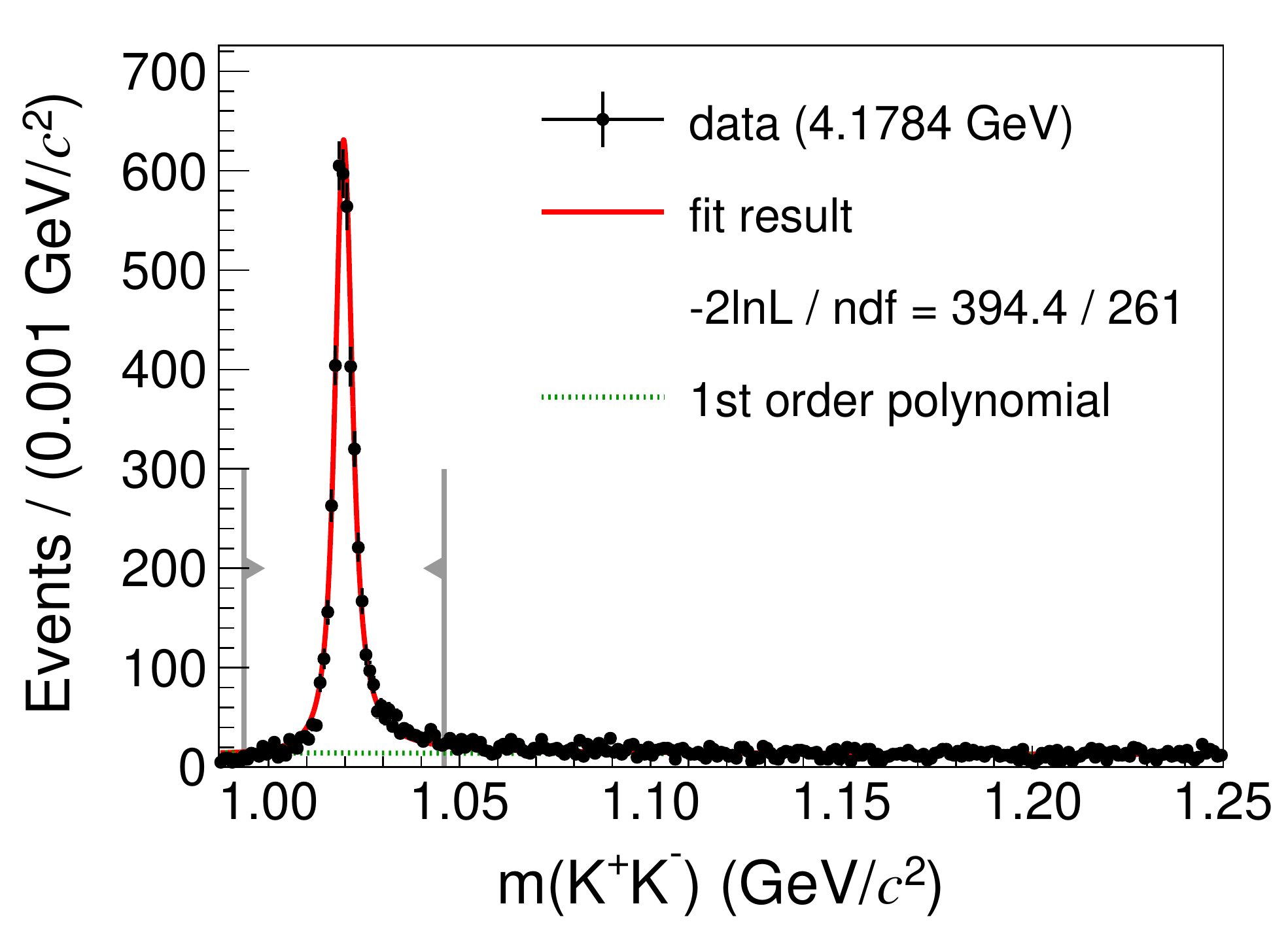}
\put(20,60){(a)}
\end{overpic}
\begin{overpic}[width=0.4\textwidth]{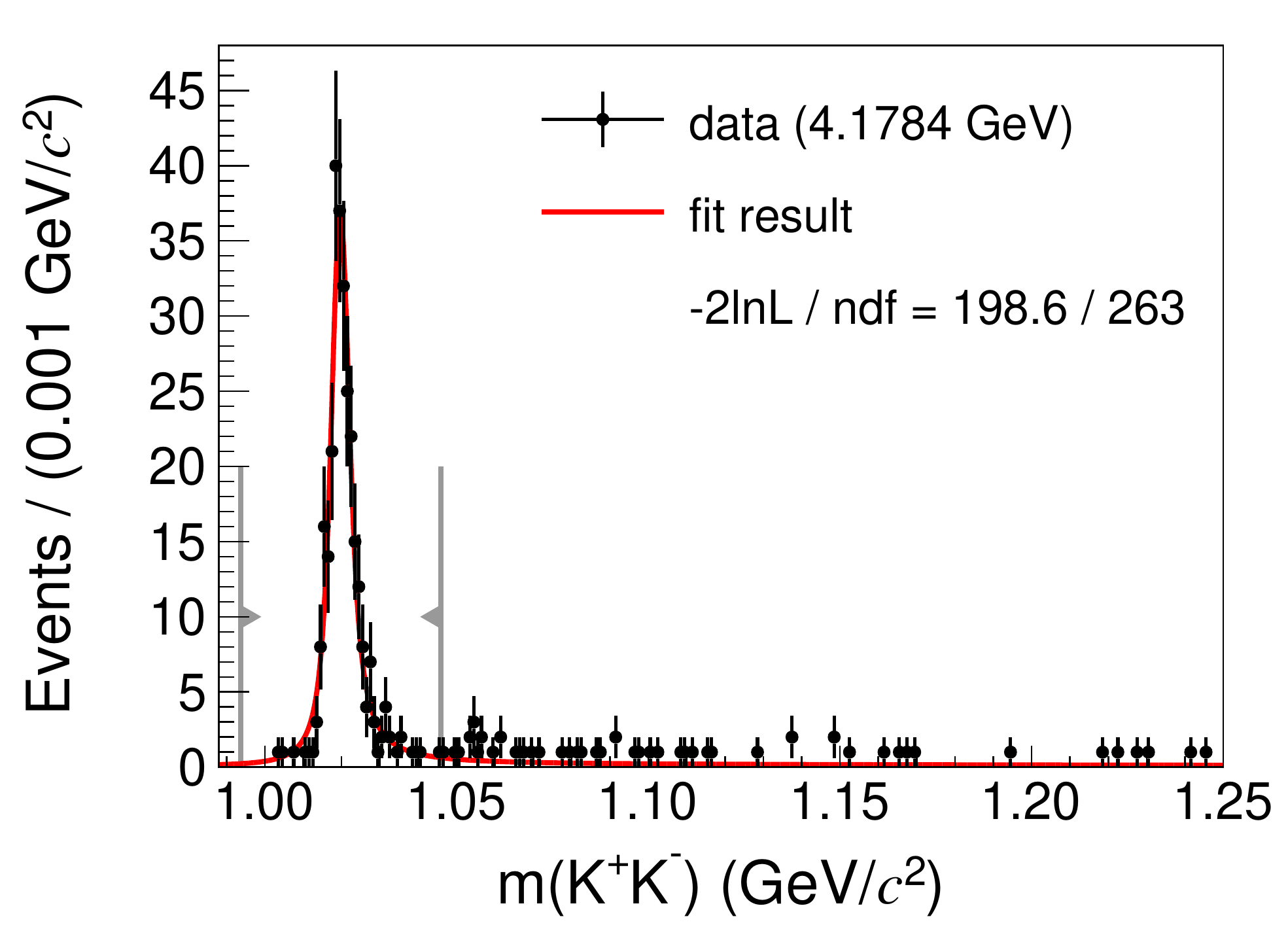}
\put(20,60){(b)}
\end{overpic}
 \caption{\label{fig:imfit} (Color online) Fits to the $\kpkm$ invariant mass distributions for the candidates of (a) $\epem\to\kpkmkpkm$ and (b) $\epem\to\kpkmksks$. Black points represent data at the center-of-mass energy of $4.1784~\gev$, full (red) curves represent the total fit result and, if relevant, short-dashed (green) curves show the background contribution. The gray markers indicate the signal region from which the number of observed events is obtained.}
\end{figure}

The final states $\kpkmkpkm$ (for $\epemtophikpkm$ with $\phi\to\kpkm$) and $\kpkm\pip\pim\pip\pim$ (for $\epemtophiksks$ with $\phi\to\kpkm$ and $\ks\to\pip\pim$) are studied in this work. 
The polar angle $\theta$ of each charged kaon track detected in the MDC has to satisfy $|\cos\theta|<0.93$, and its point of closest approach to the nominal interaction point must be within $10$~cm in the beam direction and within $1$~cm in the plane perpendicular to the beam direction. For the selection of the $\kpkm\pip\pim\pip\pim$ final state, a secondary vertex fit is performed, reconstructing two tracks of oppositely charged pions to a common vertex. It is required that the flight significance satisfy $L/\sigma_L > 2$, with flight length $L$ of $\ks$ mesons and its uncertainty $\sigma_L$.
A particle identification (PID) is performed combining the TOF and MDC information to calculate a probability $P(h)$ for the particle hypotheses $h=\pi, K, p$. The particle type with the largest probability is assigned to each track. In addition, a minimum probability of $P(h)>10^{-5}$ is required to suppress background.
A four- (six-)constraint kinematic fit is performed to the $\kpkm\kpkm$ ($\kpkm\ksks$ with $\ks\to\pip\pim$) hypothesis requiring four-momentum conservation between initial and final states and two additional mass constraints for the $\ks\to\pip\pim$ decays. The combination yielding the smallest $\chi^2$ value is used for the analysis. 
The resulting invariant mass spectra of $\phi$ meson candidates (the $\kpkm$ pair which has the closest invariant mass to the $\phi$ mass) for 
$\phi\kpkm$ and $\phi\ksks$ are displayed in Fig.~\ref{fig:imfit} for data taken at $\sqrt{s}=4.1784~\gev$.
According to the inclusive MC sample, the main background contributions are $\ee \to f_2(1270)(\to \kpkm)\kpkm$ and $\ee \to \kpkmkpkm$ (continuum production) as well as $\ee\to f_2^\prime(\to\kpkm)\ksks$, $\ee \to f_2^\prime(\to\ksks)\kpkm$ and $\ee \to \kpkm \ksks$ (continuum production) for the $\kpkmkpkm$ and $\kpkm\pip\pim\pip\pim$ final states, respectively. No peaking background is found in the $\kpkm$ invariant mass distribution in the vicinity of the $\phi$ mass.


The selection criteria with respect to the kinematic fit are optimized according to $\frac{S}{\sqrt{S+B}}$, where $S$ and $B$ are the numbers of signal and background events in the inclusive MC sample, that have been scaled to data, after requiring $\chi_\textrm{NC}^2<\chi^2_\textrm{sel}$. Here, $\chi_\textrm{NC}^2$ and $\chi^2_\textrm{sel}$ are the $\chi^2$ value of the kinematic fit with $N$ constraints and the $\chi^2$ value of the selection condition, respectively. For the identification of the $\kpkmkpkm$ final state, a $\chi_\textrm{4C}^2<93$ is found as an optimal selection condition, whereas for the $\kpkm\pipi\pipi$ final state a $\chi_\textrm{6C}^2<227$ is chosen. Since the optimum choice in $\chi_\text{NC}^2$ cuts is found to be energy-independent, the requirements on the $\chi_\textrm{NC}^2$ value are applied for all center-of-mass energies.

The number of signal events is determined from a fit to the invariant mass spectra (see Fig.~\ref{fig:imfit}).
The signal part is described by a relativistic Breit-Wigner function, taking into account the asymmetric lineshape of the $\phi$ meson due to its proximity to the $\kpkm$ threshold~\cite{asymmetric_phi}, convolved with a Gaussian function to account for the expected experimental mass resolution obtained from MC simulation. In the fit to events in the $\kpkmkpkm$ final state, the background is described by a first-order polynomial function, while no significant increase in the fit quality has been observed when introducing a background component into the corresponding fit to the $\kpkm\pipi\pipi$ final state. A binned maximum likelihood fit is performed to each dataset and final state individually, with the width of the $\phi$ meson fixed to the world average value taken from the PDG~\cite{pdg}. The number of signal events in final state $i$ ($\kpkmkpkm$ or $\kpkm\pipi\pipi$) at the center-of-mass energy $\sqrt{s}$ is
determined by integrating the signal function in the signal region. It is defined as the symmetric region around the nominal $\phi$ meson mass containing $95\%$ of all signal events according to the signal shape. Asymmetric statistical errors due to Poisson statistics for low statistics datasets are obtained via a likelihood scan of the number of signal events. These likelihood scans are parametrized by asymmetric Gaussian distributions 
\begin{eqnarray}
\label{eq:llscan}
   L_{i}(N) = \frac{1}{\sqrt{2\pi \sigma_k^2}} \cdot e^{-\frac{(N-\mu)^2}{2\sigma_k^2}} \nonumber \\
   \textrm{with}\quad \sigma_k = \left\{ \begin{matrix} \sigma_L~,~ N\leq \mu \\ \sigma_R~,~ N> \mu \end{matrix}\right.  \quad ,
\end{eqnarray}
$\mu$ being the number of observed signal events in the maximum likelihood case and $\sigma_L$ and $\sigma_R$ being the lower and upper statistical uncertainty of $\mu$ for data set $i$, respectively. Results are listed in Table~\ref{tab:xsec}.
\section{\label{sec:eff} Efficiency Determination}
The efficiency $\epsilon^i(s)$ in final state $i$ for a center-of-mass energy $\sqrt{s}$ is defined according to
%
\begin{equation} \epsilon^i(s) = \frac{N^i_\textrm{acc}(s)}{N^i_{\textrm{gen}}(s)} \, , \label{eq:acc} \end{equation}
\noindent with $N^i_{\textrm{acc}}(s)$ being the number of reconstructed signal events and $N^i_{\textrm{gen}}(s)$ being the total size of the signal MC sample. Equation~\ref{eq:acc} only provides a good representation of the efficiency if the signal MC sample properly reflects data in all relevant coordinates $\vec{x}=\{p_\phi,\theta_\phi,\varphi_\phi,p_{\kkbar},\theta_{\kkbar},\varphi_{\kkbar}, ...\}$, with $p_i, \theta_i$ and $\varphi_i$ being the radial distance, polar angle and azimuthal angle, respectively. 
Since the data distribution is not constant over the full $n$-particle phase-space and, in fact, is a priori unknown, a partial wave analysis of the data is performed in order to re-weight the MC sample.

The isobar model \cite{isobar} is used in the partial wave analysis by decomposing the full $\ee\to\gamma^*\to\phi\kpkm$ and $\ee\to\gamma^*\to\phi\ksks$ processes
into a sequence of two-body decays. Each two-body decay is described in the helicity formalism ~\cite{helicity}.
%
%
%
%
The $\ks$ meson is treated as a stable particle in the amplitude analysis. Signal MC simulations are employed to derive the line-shape of the $\phi$ meson that is used for normalization in the partial wave analysis.
Blatt-Weisskopf barrier factors~\cite{helicity} are used for both the production $\gamma^*\to a+d$ and the two-body decay $a\to b+c$ according to Ref.~\cite{maltepaper}. The final model only includes processes of the type $\ee \to \phi f_J$, with $f_J \to \kpkm$ or $f_J \to \ksks$. Due to limited statistics, we restrict ourselves to $J^{PC}=0^{++}$ and $J^{PC}=2^{++}$ quantum numbers for the $f_J$ resonances, which leads to a sufficiently good description of the data. The dynamics of the $J^{PC}=0^{++}$ contributions are described by a $K$-matrix approach up to $m(\kkbar)\leq 1.9~\gev$, incorporating the five channels $\pi\pi,K\bar{K},\eta\eta,\eta\eta^\prime$ and $4\pi$ with five fixed poles ~\cite{kmatrix}. An additional $J^{PC}=0^{++}$ resonance is included for respective states at higher invariant masses, while the $J^{PC}=2^{++}$ contributions are described by four resonances. These single resonances are parametrized as relativistic Breit-Wigner amplitudes and their masses and widths are free parameters in the fit. This is justified by the fact that the aim of this partial wave analysis is only to better describe the data so as to enable an accurate determination of the efficiency. In further model tests, no significant contribution of $\ee \to K K^*$ with $K^* \to \phi K$ is found.

The partial wave analysis is performed as an unbinned maximum likelihood fit using the software package {\sc PAWIAN}~\cite{pawian}. Details on the likelihood construction in {\sc PAWIAN} can be found in Refs.~\cite{maltepaper, pawian, malte2}. The few remaining background events underneath the $\phi$ peak ($|m_{\phi,\text{PDG}}-m(\kpkm)| < 0.01~\gevcc$) are neglected in the partial wave analysis. \\ \indent Due to the limited statistics, the data for the $\kpkm\pipi\pipi$ final state are fitted simultaneously over the whole energy range with all amplitudes fully constrained between the datasets apart from an overall scaling factor. The results of the partial wave analysis for the different final states are displayed in Fig.~\ref{pwa1} for the high statistics data taken at a center-of-mass energy of $4.1784~\gev$. For each energy point, we obtain event weights, $w(\vec{x})$, from the partial wave analysis as a function of the coordinates in the $n$-particle phase-space. The efficiency $\epsilon^i(s)$ is then determined as
\begin{equation}
 \epsilon^i(s) = \frac{\sum\limits_{j=0}^{N^i_{\textrm{acc}}(s)} w(\vec{x}_j)}{\sum\limits_{j=0}^{N^i_\textrm{gen}(s)} w(\vec{x}_j)} \, .
\end{equation}
The efficiencies obtained in this way are summarized in Table~\ref{tab:xsec}.

\begin{widetext}
$\quad$
\begin{figure}[tbh!]
\begin{overpic}[width=0.48\textwidth]{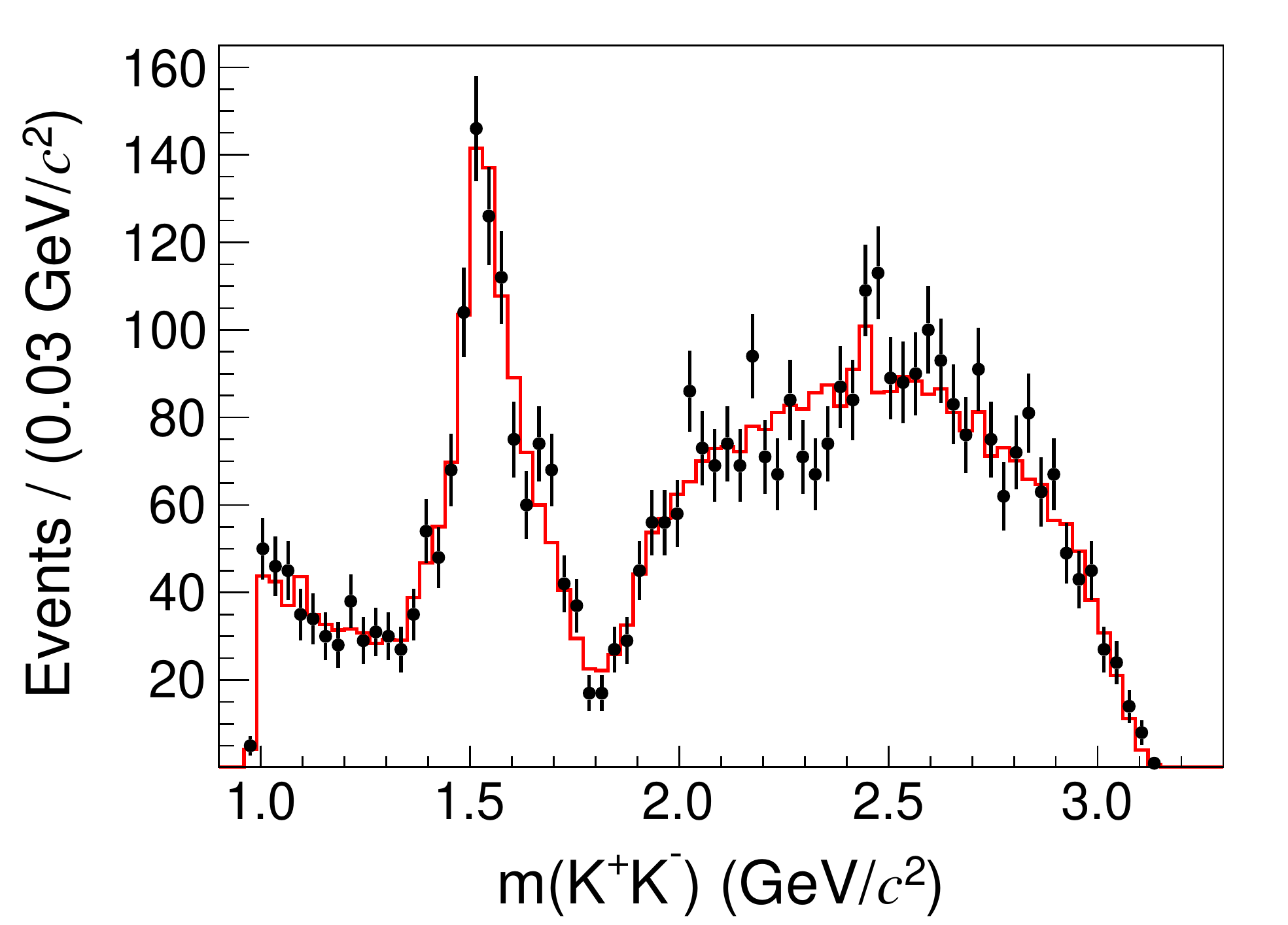}
\put(82,60){(a)}
\end{overpic}
\begin{overpic}[width=0.48\textwidth]{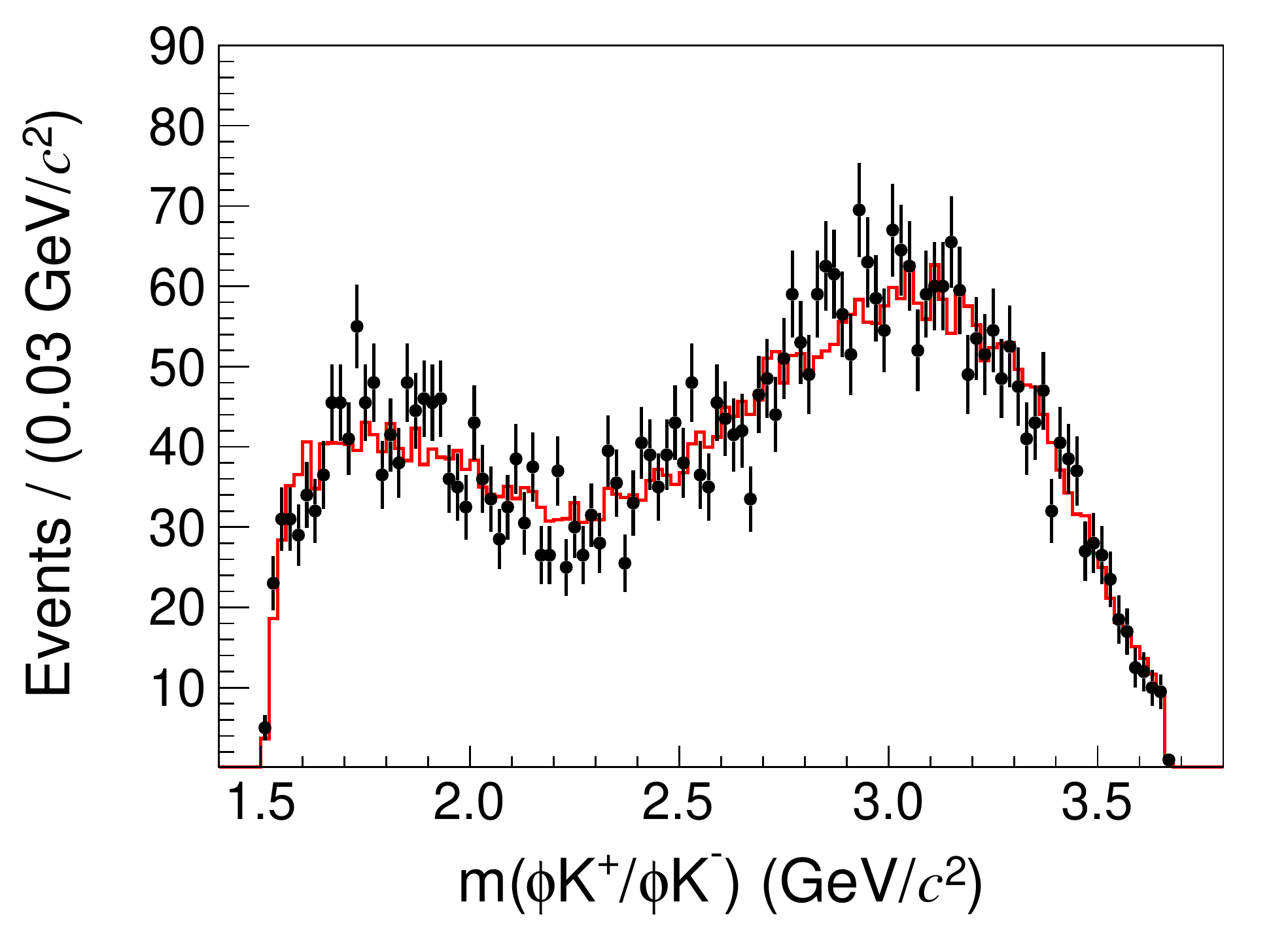}
\put(82,60){(b)}
\end{overpic}
\begin{overpic}[width=0.48\textwidth]{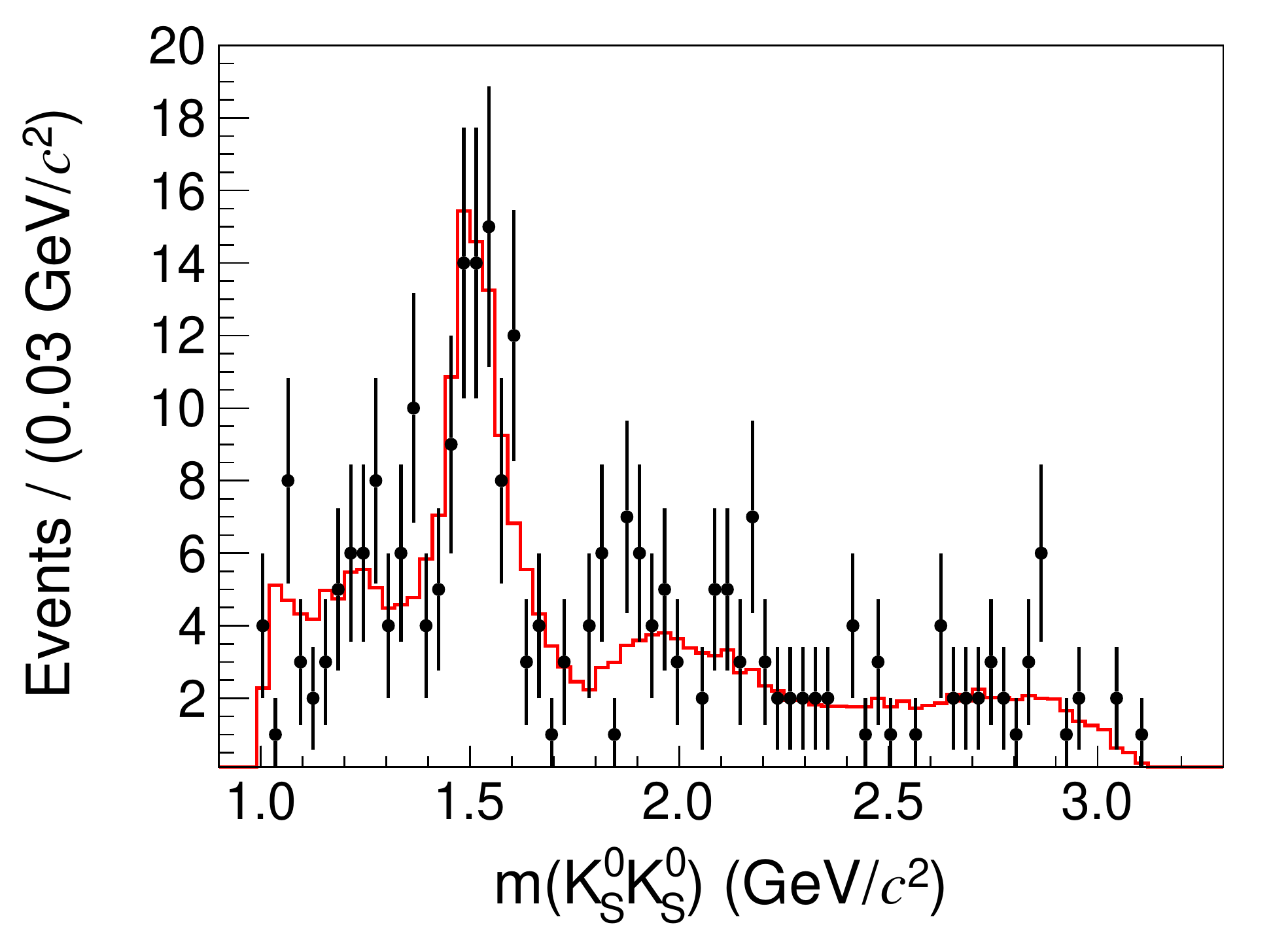}
\put(82,60){(c)}
\end{overpic}
\begin{overpic}[width=0.48\textwidth]{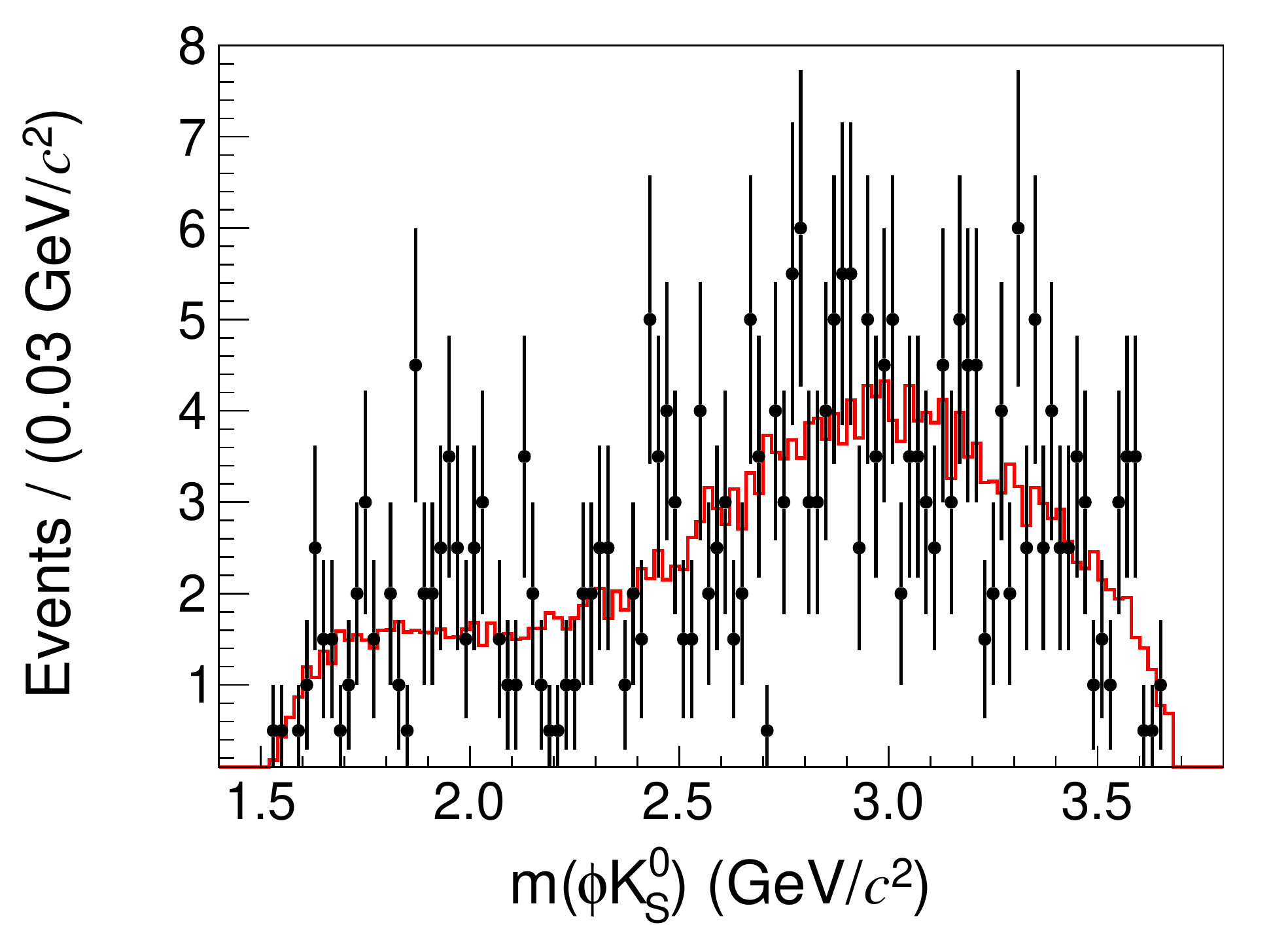}
\put(82,60){(d)}
\end{overpic}
 \caption{\label{pwa1} (Color online) Results of the partial wave analysis of the (a-b) $\epemtophikpkm$ and (c-d) $\epemtophiksks$ processes for the data taken at a center-of-mass energy of $4.1784~\gev$. The left column shows the invariant mass of the $\kpkm(\ksks)$ system recoiling off the $\phi$ meson, the right column the invariant mass of the $\phi\kp/\phi\km$ ($\phi\ks$) system. Black points correspond to data and full (red) curves show the result of the amplitude analysis.}
\end{figure}
\end{widetext}

%
%



\section{\label{sec:born} Determination of Born Cross Sections}

The Born cross sections of the processes $\epemtophikpkm$ and $\epemtophiksks$ are determined by
\begin{equation}
    \sigma_B(s) = \frac{N(s)}{L(s)\cdot (1+\delta_r(s))\cdot \frac{1}{|1-\Pi|^2} \cdot \epsilon(s)\cdot Br} \, , \label{eq:born}
\end{equation}
with $N(s)$ denoting the number of signal events observed in the data at the center-of-mass energy $\sqrt{s}$, $L(s)$ being the corresponding integrated luminosity determined using Bhabha scattering~\cite{lumi}, $\delta_r(s)$ and $\frac{1}{|1-\Pi|^2}$ being corrections accounting for ISR and VP, $\epsilon(s)$ being the efficiency and $Br$ corresponding to the product of branching ratios involved in the decay ($Br=Br(\phi\to\kpkm)$ for $\epemtophikpkm$ and $Br=Br(\phi\to\kpkm)\,\times\, Br(\ks\to\pipi)^2$ for $\epemtophiksks$).
The correction $\frac{1}{|1-\Pi|^2}$ is calculated with the {\sc alphaQED} software package~\cite{alphaqed} with an accuracy of $0.5\%$. The ISR effect is treated in an iterative procedure starting from a flat energy dependence of the Born cross section $\sigma_B(s)$. It depends on the shape of the cross section and can, in general, have an effect on the efficiency. The procedure is considered converged once two successive iterations $i$ and $i-1$ give $\kappa_i/\kappa_{i-1}=1$ within statistical uncertainties, where $\kappa(s)=\epsilon(s)\cdot (1+\delta_r(s))$ is the product of the efficiency and the corresponding radiative correction factor $1+\delta_r(s)$ obtained from the {\sc KKMC} MC generator for each iteration~\cite{ISR_procedure}.
%
%
%
%
The resulting Born cross sections are shown in Fig.~\ref{fig:xsec}. Table~\ref{tab:xsec} summarizes the Born cross sections together with the relevant values that are used for the calculation.

\begin{widetext}
$\quad$
{\renewcommand{\arraystretch}{1.3}%
\begin{table}[h]
 \caption{\label{tab:xsec} Summary of the Born cross sections $\sigma_B$ of the processes $\epemtophikpkm$ ($\sigma_{B_1}$) and $\epemtophiksks$ ($\sigma_{B_2}$) for the datasets at different center-of-mass energies $\sqrt{s}$, integrated luminosity $L$, number of observed events $N_{1,2}$, efficiency $\epsilon_{1,2}$, radiative corrections $(1+\delta_r)_{1,2}$ and the vacuum polarization correction $\frac{1}{|1-\Pi|^2}$.}
 \begin{tabular}{c|c|c|c|c|c|c|c|c|c|c}
  \hline
  $\sqrt{s}$ (GeV) & $L$ (pb$^{-1}$) & $N_1$ & $\varepsilon_1$ (\%) & $(1+\delta_r)_1$ & $N_2$ & $\varepsilon_2$ (\%) & $(1+\delta_r)_2$ & $\frac{1}{|1-\Pi|^2}$ & $\sigma_{B_1}$ (pb)  & $\sigma_{B_2}$ (pb)\\
  \hline
		$3.7730$ & $2931.8$ & $7329.2^{+94.6}_{-92.9}$ & $35.8 \pm 0.2$ & $0.8295 $ & $360.8^{+20.9}_{-19.2}$ & $18.0 \pm 0.1$ & $0.8281 $ & $1.0560$ & $16.19^{+0.24}_{-0.23} \pm 0.73$ & $3.31^{+0.19}_{-0.18} \pm 0.10$ \\ [1pt]
		$3.8695$ & $224.0$ & $452.5^{+24.1}_{-22.4}$ & $34.7 \pm 0.2$ & $0.8889 $ & $27.1^{+5.9}_{-4.3}$ & $17.6 \pm 0.1$ & $0.8864 $ & $1.0506$ & $12.68^{+0.68}_{-0.63} \pm 0.61$ & $3.14^{+0.69}_{-0.50} \pm 0.10$ \\ [1pt]
		$4.0076$ & $482.0$ & $825.3^{+32.5}_{-30.8}$ & $33.7 \pm 0.2$ & $0.9390 $ & $56.6^{+8.2}_{-6.6}$ & $16.2 \pm 0.2$ & $0.9346 $ & $1.0441$ & $10.53^{+0.43}_{-0.41} \pm 0.47$ & $3.16^{+0.46}_{-0.37} \pm 0.09$ \\ [1pt]
		$4.1285$ & $401.5$ & $590.3^{+27.4}_{-25.7}$ & $32.4 \pm 0.2$ & $0.9675 $ & $34.7^{+6.6}_{-5.0}$ & $15.8 \pm 0.1$ & $0.9611 $ & $1.0525$ & $9.06^{+0.43}_{-0.41} \pm 0.38$ & $2.30^{+0.44}_{-0.33} \pm 0.07$ \\ [1pt]
		$4.1574$ & $408.7$ & $633.9^{+28.4}_{-26.7}$ & $32.4 \pm 0.2$ & $0.9739 $ & $27.1^{+5.9}_{-4.3}$ & $15.2 \pm 0.1$ & $0.9667 $ & $1.0533$ & $9.48^{+0.44}_{-0.41} \pm 0.39$ & $1.82^{+0.40}_{-0.29} \pm 0.05$ \\ [1pt]
		$4.1784$ & $3189.0$ & $4572.5^{+74.9}_{-73.2}$ & $32.6 \pm 0.2$ & $0.9783 $ & $289.6^{+18.6}_{-16.9}$ & $15.9 \pm 0.1$ & $0.9710 $ & $1.0541$ & $8.67^{+0.17}_{-0.17} \pm 0.36$ & $2.37^{+0.16}_{-0.14} \pm 0.07$ \\ [1pt]
		$4.1888$ & $524.6$ & $754.9^{+30.9}_{-29.2}$ & $32.6 \pm 0.2$ & $0.9801 $ & $49.9^{+7.7}_{-6.1}$ & $16.1 \pm 0.2$ & $0.9727 $ & $1.0558$ & $8.68^{+0.37}_{-0.35} \pm 0.36$ & $2.44^{+0.38}_{-0.30} \pm 0.07$ \\ [1pt]
		$4.1989$ & $526.0$ & $785.2^{+31.6}_{-29.9}$ & $32.3 \pm 0.2$ & $0.9823 $ & $42.3^{+7.2}_{-5.6}$ & $15.9 \pm 0.1$ & $0.9746 $ & $1.0564$ & $9.05^{+0.38}_{-0.36} \pm 0.38$ & $2.09^{+0.35}_{-0.28} \pm 0.06$ \\ [1pt]
		$4.2091$ & $518.0$ & $694.2^{+29.6}_{-27.9}$ & $31.6 \pm 0.2$ & $0.9846 $ & $47.1^{+7.5}_{-5.9}$ & $16.2 \pm 0.1$ & $0.9765 $ & $1.0568$ & $8.28^{+0.37}_{-0.35} \pm 0.34$ & $2.31^{+0.37}_{-0.29} \pm 0.07$ \\ [1pt]
		$4.2187$ & $514.6$ & $673.6^{+29.2}_{-27.5}$ & $32.1 \pm 0.2$ & $0.9866 $ & $48.0^{+7.6}_{-6.0}$ & $16.0 \pm 0.1$ & $0.9784 $ & $1.0563$ & $7.94^{+0.36}_{-0.34} \pm 0.33$ & $2.39^{+0.38}_{-0.30} \pm 0.07$ \\ [1pt]
		$4.2263$ & $1056.4$ & $1390.2^{+41.6}_{-39.9}$ & $33.1 \pm 0.2$ & $0.9882 $ & $79.4^{+9.5}_{-7.9}$ & $16.0 \pm 0.1$ & $0.9798 $ & $1.0564$ & $7.74^{+0.25}_{-0.24} \pm 0.32$ & $1.92^{+0.23}_{-0.19} \pm 0.06$ \\ [1pt]
		$4.2357$ & $530.3$ & $715.6^{+30.1}_{-28.4}$ & $32.2 \pm 0.2$ & $0.9902 $ & $51.8^{+7.8}_{-6.3}$ & $16.5 \pm 0.2$ & $0.9820 $ & $1.0555$ & $8.14^{+0.35}_{-0.34} \pm 0.34$ & $2.43^{+0.37}_{-0.30} \pm 0.07$ \\ [1pt]
		$4.2438$ & $538.1$ & $659.4^{+29.4}_{-27.7}$ & $32.1 \pm 0.2$ & $0.9919 $ & $35.7^{+6.7}_{-5.1}$ & $15.8 \pm 0.1$ & $0.9832 $ & $1.0555$ & $7.42^{+0.34}_{-0.32} \pm 0.31$ & $1.72^{+0.32}_{-0.25} \pm 0.05$ \\ [1pt]
		$4.2580$ & $828.4$ & $978.9^{+35.4}_{-33.6}$ & $32.9 \pm 0.2$ & $0.9951 $ & $69.9^{+9.0}_{-7.4}$ & $16.1 \pm 0.1$ & $0.9867 $ & $1.0536$ & $6.96^{+0.26}_{-0.25} \pm 0.29$ & $2.14^{+0.28}_{-0.23} \pm 0.06$ \\ [1pt]
		$4.2666$ & $531.1$ & $697.3^{+29.7}_{-28.0}$ & $32.7 \pm 0.2$ & $0.9970 $ & $33.8^{+6.5}_{-4.9}$ & $16.0 \pm 0.1$ & $0.9882 $ & $1.0532$ & $7.76^{+0.34}_{-0.32} \pm 0.32$ & $1.62^{+0.31}_{-0.24} \pm 0.05$ \\ [1pt]
		$4.2776$ & $175.7$ & $241.6^{+17.5}_{-15.8}$ & $31.4 \pm 0.2$ & $0.9990 $ & $13.8^{+4.5}_{-2.9}$ & $17.0 \pm 0.2$ & $0.9898 $ & $1.0530$ & $8.47^{+0.62}_{-0.56} \pm 0.36$ & $1.88^{+0.61}_{-0.40} \pm 0.06$ \\ [1pt]
		$4.2879$ & $502.4$ & $577.7^{+27.2}_{-25.5}$ & $31.4 \pm 0.2$ & $1.0010 $ & $37.6^{+6.8}_{-5.2}$ & $15.7 \pm 0.1$ & $0.9918 $ & $1.0527$ & $7.07^{+0.34}_{-0.32} \pm 0.30$ & $1.94^{+0.35}_{-0.27} \pm 0.06$ \\ [1pt]
		$4.3121$ & $501.2$ & $583.1^{+27.5}_{-25.7}$ & $31.9 \pm 0.2$ & $1.0056 $ & $35.7^{+6.7}_{-5.1}$ & $15.8 \pm 0.1$ & $0.9955 $ & $1.0522$ & $7.00^{+0.34}_{-0.32} \pm 0.30$ & $1.82^{+0.34}_{-0.26} \pm 0.06$ \\ [1pt]
		$4.3374$ & $505.0$ & $556.2^{+26.5}_{-24.8}$ & $31.0 \pm 0.2$ & $1.0105 $ & $25.2^{+5.7}_{-4.2}$ & $14.9 \pm 0.1$ & $1.0001 $ & $1.0508$ & $6.81^{+0.33}_{-0.31} \pm 0.29$ & $1.36^{+0.31}_{-0.22} \pm 0.04$ \\ [1pt]
		$4.3583$ & $543.9$ & $604.6^{+27.6}_{-25.9}$ & $33.1 \pm 0.2$ & $1.0141 $ & $45.2^{+7.4}_{-5.8}$ & $16.7 \pm 0.2$ & $1.0032 $ & $1.0511$ & $6.41^{+0.30}_{-0.28} \pm 0.27$ & $2.01^{+0.33}_{-0.26} \pm 0.06$ \\ [1pt]
		$4.3774$ & $522.7$ & $593.3^{+27.5}_{-25.8}$ & $32.4 \pm 0.2$ & $1.0177 $ & $41.4^{+7.1}_{-5.5}$ & $16.1 \pm 0.1$ & $1.0063 $ & $1.0513$ & $6.65^{+0.32}_{-0.30} \pm 0.28$ & $1.98^{+0.34}_{-0.27} \pm 0.06$ \\ [1pt]
		$4.3964$ & $507.8$ & $530.2^{+26.1}_{-24.4}$ & $32.0 \pm 0.2$ & $1.0214 $ & $33.8^{+6.5}_{-4.9}$ & $16.0 \pm 0.1$ & $1.0094 $ & $1.0510$ & $6.17^{+0.31}_{-0.29} \pm 0.26$ & $1.67^{+0.32}_{-0.24} \pm 0.05$ \\ [1pt]
		$4.4156$ & $1043.9$ & $1114.1^{+37.3}_{-35.6}$ & $32.4 \pm 0.2$ & $1.0244 $ & $75.6^{+9.3}_{-7.7}$ & $16.7 \pm 0.1$ & $1.0118 $ & $1.0524$ & $6.21^{+0.22}_{-0.21} \pm 0.26$ & $1.72^{+0.21}_{-0.18} \pm 0.05$ \\ [1pt]
		$4.4362$ & $569.9$ & $619.2^{+28.2}_{-26.5}$ & $32.5 \pm 0.2$ & $1.0275 $ & $32.8^{+6.4}_{-4.8}$ & $15.6 \pm 0.1$ & $1.0147 $ & $1.0537$ & $6.27^{+0.29}_{-0.28} \pm 0.27$ & $1.47^{+0.29}_{-0.22} \pm 0.04$ \\ [1pt]
		$4.4671$ & $111.1$ & $93.4^{+11.6}_{-9.9}$ & $32.6 \pm 0.2$ & $1.0325 $ & $8.1^{+3.7}_{-2.1}$ & $18.7 \pm 0.2$ & $1.0193 $ & $1.0548$ & $4.81^{+0.60}_{-0.51} \pm 0.22$ & $1.55^{+0.69}_{-0.40} \pm 0.05$ \\ [1pt]
		$4.5271$ & $112.1$ & $97.8^{+11.5}_{-9.8}$ & $33.2 \pm 0.2$ & $1.0427 $ & $3.4^{+2.7}_{-1.2}$ & $18.0 \pm 0.2$ & $1.0282 $ & $1.0545$ & $4.86^{+0.57}_{-0.49} \pm 0.23$ & $0.66^{+0.52}_{-0.22} \pm 0.02$ \\ [1pt]
		$4.5995$ & $586.9$ & $533.4^{+26.0}_{-24.3}$ & $32.2 \pm 0.2$ & $1.0543 $ & $30.0^{+6.2}_{-4.6}$ & $16.5 \pm 0.1$ & $1.0382 $ & $1.0546$ & $5.16^{+0.26}_{-0.24} \pm 0.22$ & $1.20^{+0.25}_{-0.18} \pm 0.04$ \\ [1pt]
		$4.6151$ & $102.5$ & $69.3^{+9.9}_{-8.2}$ & $32.7 \pm 0.2$ & $1.0569 $ & $4.3^{+2.9}_{-1.4}$ & $16.9 \pm 0.1$ & $1.0403 $ & $1.0545$ & $3.77^{+0.54}_{-0.45} \pm 0.18$ & $0.97^{+0.65}_{-0.31} \pm 0.03$ \\ [1pt]
		$4.6304$ & $511.1$ & $424.3^{+23.5}_{-21.8}$ & $32.4 \pm 0.2$ & $1.0592 $ & $21.4^{+5.4}_{-3.8}$ & $15.5 \pm 0.1$ & $1.0424 $ & $1.0544$ & $4.66^{+0.26}_{-0.24} \pm 0.20$ & $1.04^{+0.26}_{-0.18} \pm 0.03$ \\ [1pt]
		$4.6431$ & $541.4$ & $407.0^{+23.0}_{-21.3}$ & $31.9 \pm 0.2$ & $1.0612 $ & $19.5^{+5.2}_{-3.6}$ & $16.0 \pm 0.1$ & $1.0440 $ & $1.0544$ & $4.28^{+0.24}_{-0.23} \pm 0.18$ & $0.87^{+0.23}_{-0.16} \pm 0.03$ \\ [1pt]
		$4.6639$ & $523.6$ & $398.5^{+22.7}_{-21.0}$ & $31.9 \pm 0.2$ & $1.0644 $ & $31.9^{+6.3}_{-4.8}$ & $15.6 \pm 0.1$ & $1.0466 $ & $1.0544$ & $4.32^{+0.25}_{-0.23} \pm 0.18$ & $1.50^{+0.30}_{-0.23} \pm 0.05$ \\ [1pt]
		$4.6842$ & $1631.7$ & $1295.0^{+40.5}_{-38.7}$ & $32.0 \pm 0.2$ & $1.0677 $ & $61.3^{+8.5}_{-6.9}$ & $15.3 \pm 0.1$ & $1.0492 $ & $1.0545$ & $4.48^{+0.14}_{-0.14} \pm 0.19$ & $0.94^{+0.13}_{-0.11} \pm 0.03$ \\ [1pt]
		$4.7008$ & $526.2$ & $389.6^{+22.7}_{-21.0}$ & $31.9 \pm 0.2$ & $1.0704 $ & $22.4^{+5.5}_{-3.9}$ & $14.0 \pm 0.1$ & $1.0515 $ & $1.0545$ & $4.18^{+0.24}_{-0.23} \pm 0.18$ & $1.16^{+0.28}_{-0.20} \pm 0.04$ \\ [1pt]
  \hline
  \end{tabular}
 \end{table}}
 
 \end{widetext}

\section{Systematic Uncertainties}

The integrated luminosity has been determined using Bhabha scattering and its uncertainty is found to be $1\%$ \cite{lumi}. The systematic uncertainty of the tracking efficiency has been determined using a $\jpsi\to\ppbar\pipi$ control sample~\cite{tracking} and estimated to be $1\%$ per track. For multiple final state charged particles per event, the corresponding uncertainties of each track are added linearly. 
Uncertainties on the quoted branching fractions are taken from the PDG~\cite{pdg}.

Additional uncertainties due to selection conditions are investigated by varying this selection condition around its default value. The resulting Born cross section is determined and compared with the nominal value using the ratio $R=\frac{\sigma_\textrm{step}}{\sigma_\textrm{nom}}$. The systematic uncertainty is estimated as a standard deviation of a weighted sample of $R$. Here, $1/\delta R$ is used as the weight, where $\delta R$ is the uncertainty taking the sizeable correlation between the event samples into account. With regard to the kinematic fit, where a selection condition of $\chi^2<93$ ($\chi^2<227$) is applied in case of the $\kpkmkpkm$ ($\kpkm\pipi\pipi$) final state, the selection condition is varied between $\chi^2<43$ ($\chi^2<177$) and $\chi^2<143$ ($\chi^2<277$) in steps of $\delta\chi^2=5$. 
For the background description, the polynomial shapes are increased by one and two orders from the nominal first-order polynomial used to fit the $\kpkm$ invariant mass spectrum in case of the $\kpkmkpkm$ final state. 
For the ISR correction factor, we perform two additional iterations and found no significant difference and therefore neglected it as a source for systematic uncertainty.
The uncertainty associated with the $\ks$ reconstruction is determined based on studies of the control samples $J/\psi\to K^*(892)^\mp K^\pm$ and $J/\psi \to \phi \ks K^\mp \pi^\pm$~\cite{ks_sys}.
The nominal symmetric signal region containing $95\%$ of the total signal is altered to a set of both smaller and larger signal regions. No systematic effect is observed here.
To estimate a systematic uncertainty arising from the choice of the PWA model, an additional $J^{PC}=0^{++}$ as well as $J^{PC}=2^{++}$ resonance is added. The deviation in the efficiency and hence Born cross section is used as the systematic uncertainty. Its value is found to be energy-independent.
The systematic uncertainties are summarized in Table~\ref{tab:sys} for the data taken at a center-of-mass energy of $4.1784~\gev$. The total systematic uncertainty is obtained by adding each contribution in quadrature. The systematic uncertainties are obtained for various data sets individually.

{\renewcommand{\arraystretch}{1.0}%
\begin{table}
 \caption{\label{tab:sys} Summary of relative systematic uncertainties of the Born cross section measurement in percent for the data at $\sqrt{s}=4.1784~\gev$.}
 \begin{tabular}{l|c|c}
  \hline
   & $\kpkmkpkm$ & $\kpkm\pipi\pipi$ \\
  \hline
  Luminosity                & 1.0 & 1.0  \\
  Tracking efficiency       & 4.0 & 2.0  \\
  Branching fraction        & 1.0 & 1.0  \\
  $\chisq$ cut              & 0.1 &  0.2       \\
  Background description    & 0.4 &   \\
   $\ks$ reconstruction    &  &  2.0 \\
   PWA model				& 1.0 & 1.0 \\
  \hline
  Total                     & 4.4 & 3.3  \\
  \hline
  \end{tabular}
 \end{table}

\section{Search For Resonant Contributions}

In order to search for possible  $\ee\to V\to\phi\kpkm$ ($\ee\to V\to\phi\ksks$) resonant contributions, two different fits are performed. In the first fit, only a non-resonant contribution of the type~\cite{ref7}
\begin{equation}
    \sigma_{\textrm{nr}}(s) = \left(\frac{C}{\sqrt{s}}\right)^\lambda 
\end{equation}
is used. The second fit includes a single Breit-Wigner amplitude of the form~\cite{breitwigner}
\begin{equation}
    A_\textrm{res}(s) =  
	\frac{\sqrt{12\pi \Gamma_{\epem} Br(V\to\phikkbar)\Gamma}}{s-m^2+im\Gamma}
\end{equation}
that is coherently added to the non-resonant term (with $\hbar^2c^2/\SI{}{GeV^{2}}=\SI{0.3894}{mb}$), where $m$ and $\Gamma$ denote the mass and width of the resonance, respectively. The product $\zeta_V=\Gamma_{\epem}Br(V\to\phikkbar)$ of the electronic width $\Gamma_{\epem}$ and the branching fraction $Br(V\to\phikkbar)$  of the resonance $V$ is a free parameter in the fit and is associated with the amplitudes strength. 

\begin{figure}[htb!]
\begin{overpic}[width=0.45\textwidth]{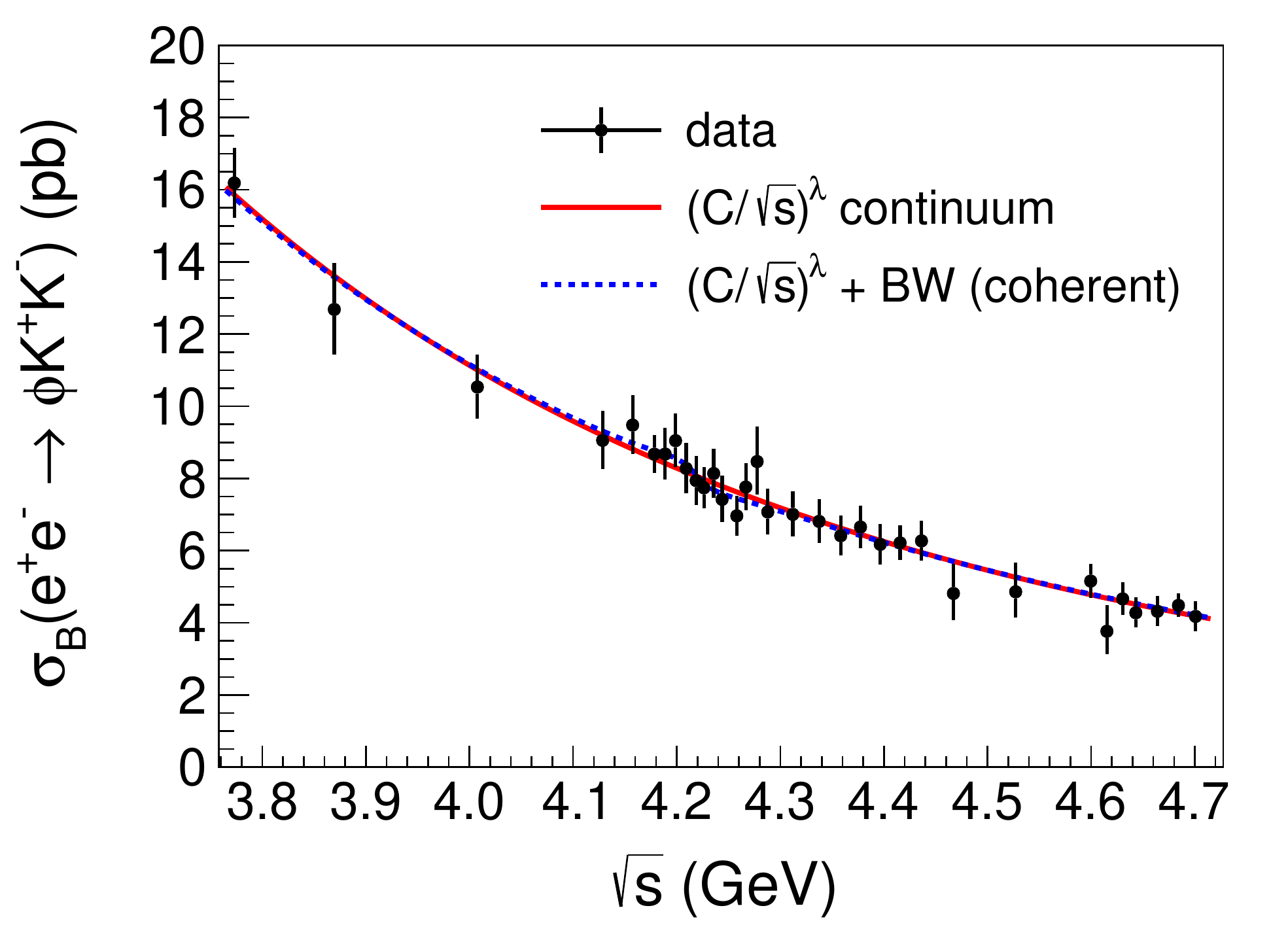}
\put(35,60){(a)}
\end{overpic}
\begin{overpic}[width=0.45\textwidth]{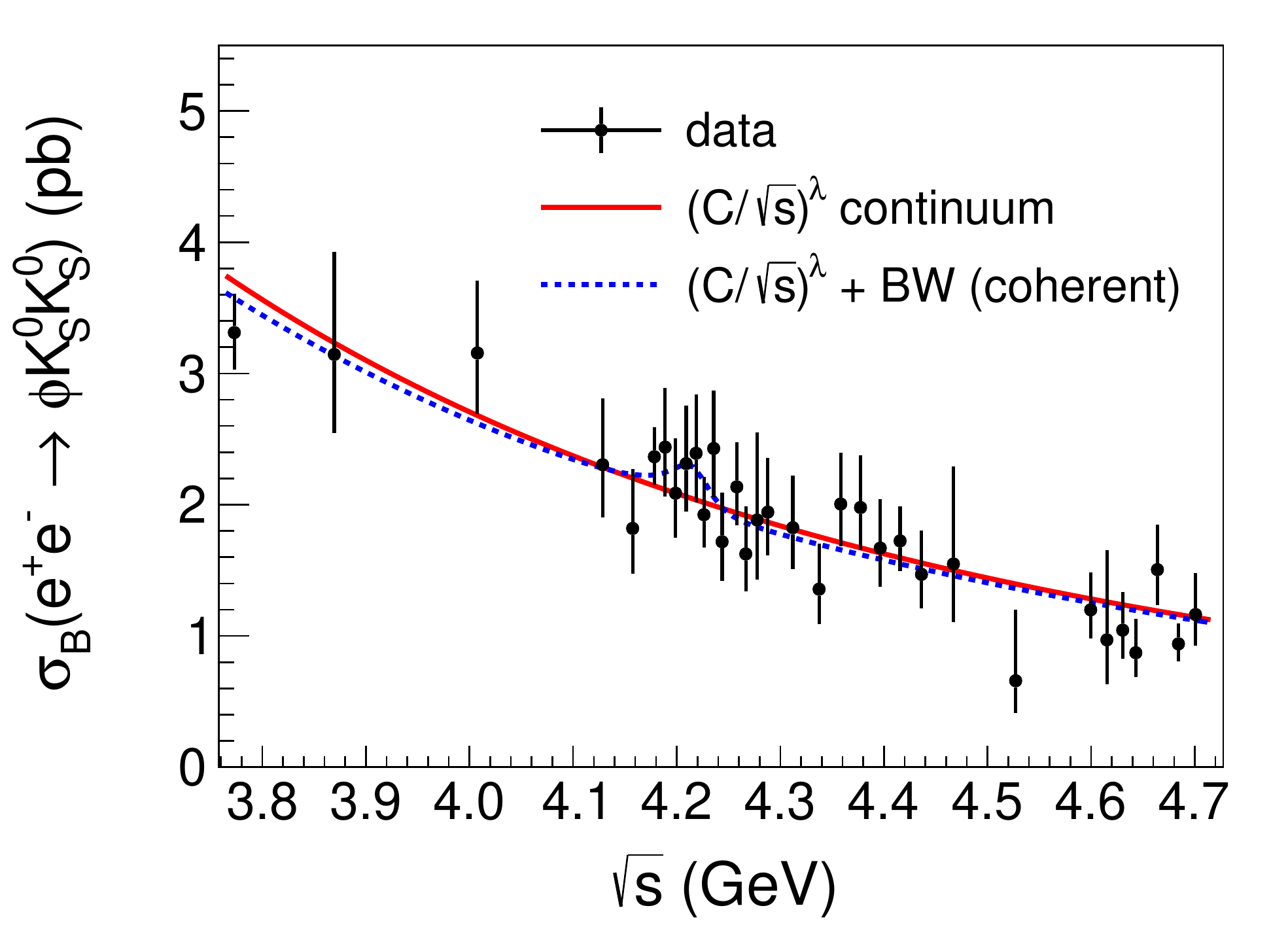}
\put(35,60){(b)}
\end{overpic}
\caption{\label{fig:xsec} (Color online) Born cross sections of the (a) $\epemtophikpkm$ and (b) $\epemtophiksks$ processes as a function of the center-of-mass energy. Black points represent our results including both statistical and systematic uncertainties. The full (red) and dashed (blue) 
curves represent the fits using a continuum contribution (with \mbox{$C = (5.95\pm0.09)\,\text{GeV}\,\text{pb}^{-\lambda}$}, \mbox{$\lambda = 6.06\pm0.28$} for \mbox{$\epemtophikpkm$} and \mbox{$C = (4.82\pm0.06)\,\text{GeV}\,\text{pb}^{-\lambda}$}, \mbox{$\lambda = 5.35\pm0.47$} for \mbox{$\epemtophiksks$}) and a Breit-Wigner coherently added to the continuum contribution, respectively. The fits displayed ($m=4.2187~\gevcc$ and $\Gamma=44~\mev$) are those for the current world average parameters of the $\psi(4230)$~\cite{pdg}.}
\end{figure}

Maximum likelihood fits are performed where the likelihood $L(x;\Theta)$ with the given data $x$ and the fit parameters $\Theta$ is defined as the product $L(x;\Theta)=\prod_{i}L_{i}(\Theta)$, with $L_{i}$ being the likelihood function for dataset $i$. 
These likelihood functions are transformed such that they only depend on the expected number of signal events $N\equiv N_{i}(\Theta)$ which can be calculated for each dataset according to Eq.~\ref{eq:born}. The likelihoods $L_{i}(N)$ obtained from data in Eq.~\ref{eq:llscan} are modified to incorporate the systematic uncertainties of dataset $i$.
%
%
In the fit, all systematic uncertainties apart from the one on the branching ratio of the meson decays are considered uncorrelated between the different center-of-mass energies. While a correlation of a systematic uncertainty between two center-of-mass energies cannot in general be ruled out, our assumption of a vanishing correlation leads to the most conservative signal estimate.

No evidence for a resonant contribution from the fits is found
and the upper limit for a wide range of resonance parameters $m$ and $\Gamma$ at the $90\%$ confidence level are set. As the resonant contribution is added coherently according to 
\begin{equation}
\sigma_\textrm{coh} = \left | \sqrt{\sigma_\textrm{nr}(s)} + A_\textrm{res}(s) \cdot e^{i\phi}\right|^2 \, 
\end{equation}
with phase $\phi$, the fit finds two ambiguous solutions for constructive and destructive interference~\cite{kai}. The upper limits are obtained by integrating $L(x;\Theta)=\prod_{i}L_{i}(\Theta)$ according to
\begin{equation}
\frac{\int\limits_{-\infty}^{\zeta_V^{\textrm{UL}}} L(x,\Theta) \, \pi(\Theta) \,  d\zeta_V}{\int\limits_{-\infty}^{\infty} L(x,\Theta) \, \pi(\Theta) \, d\zeta_V} = 0.90 ~ ,
\end{equation}
where the prior $\pi(\Theta)$ is given by
\begin{equation}
\pi(\Theta) = \left\{ \begin{matrix} 1 ~ , ~ \zeta_V \geq 0 \\ 0 ~ , ~ \zeta_V < 0 \end{matrix} \right. \quad .
\end{equation}
The procedure outlined above is repeated  with a step size of $1~\mev$ for different masses $m$ in the range $4.15~\gevcc < m < 4.45~\gevcc$ and widths $\Gamma$ in the range $40~\mev < \Gamma < 240~\mev$ for a potential resonant contribution. The results are shown in Fig.~\ref{fig:ul}.
%

\begin{figure}[htb]
 \includegraphics[width=0.45\textwidth]{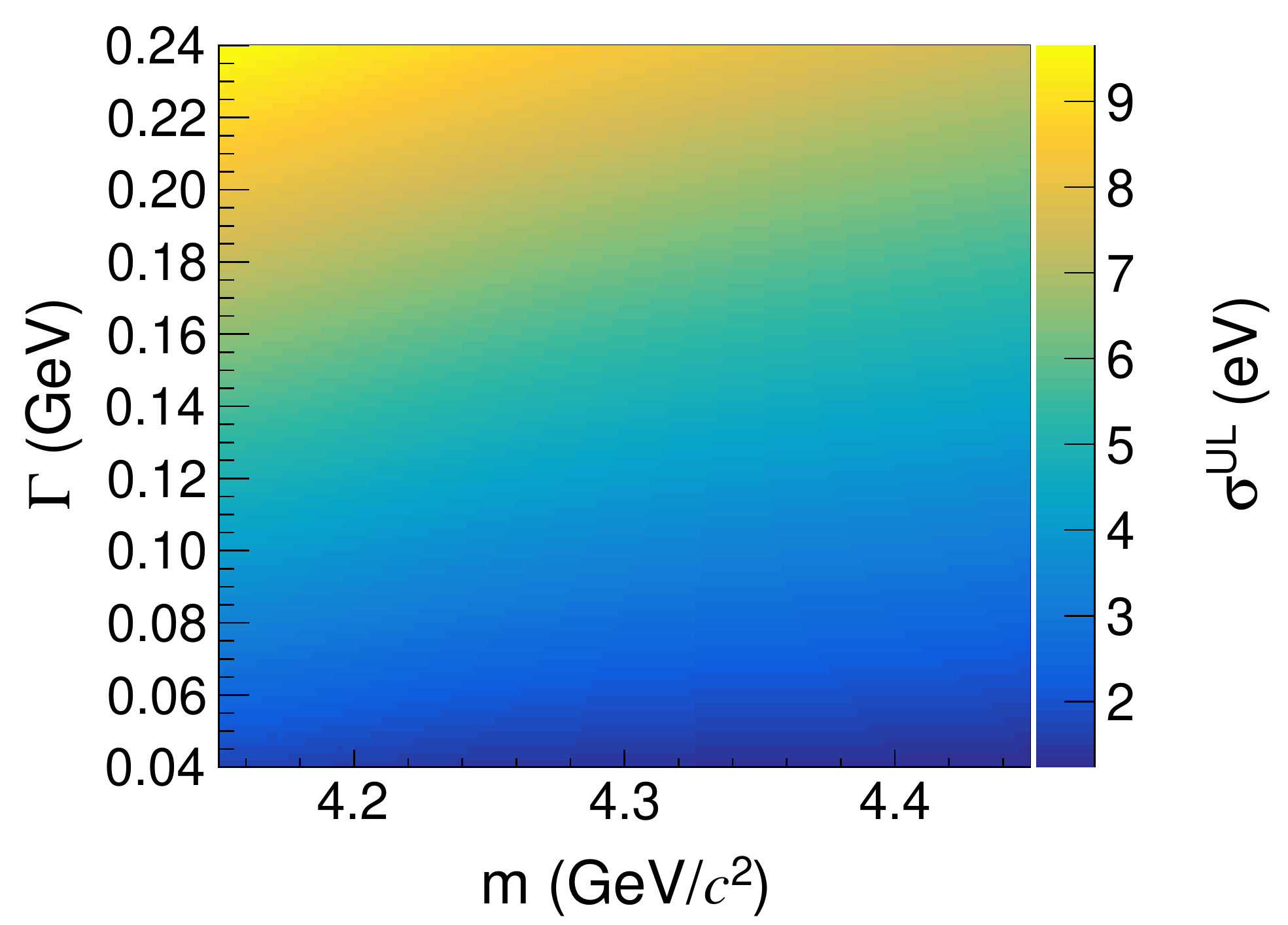}
 \includegraphics[width=0.45\textwidth]{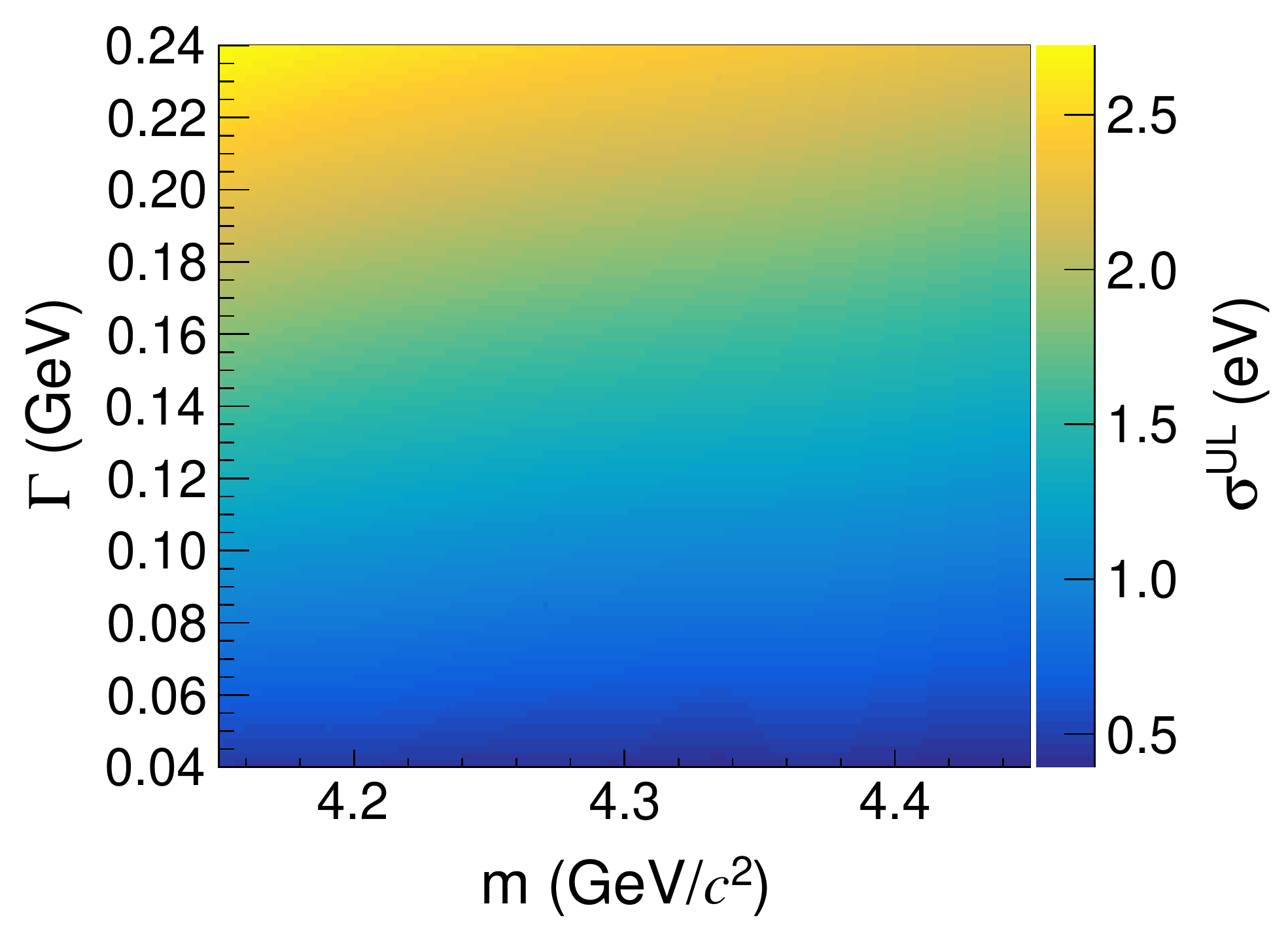}
 \caption{\label{fig:ul} (Color online) Upper limits on a possible resonant contribution with mass $m$ and width $\Gamma$ added coherently to the continuum contribution
 for $\epemtophikpkm$ (top) and $\epemtophiksks$ (bottom).}
\end{figure}
%
The dependence of the upper limits on the mass and widths of a resonant contribution $V$ is observed to be flat.
The upper limits for a resonant contribution of the $\psi(4230)$, using current world average values for mass ($m=4.2187~\gevcc$) and width ($\Gamma=44~\mev$) \cite{pdg} are $\zeta_{V,\text{coh}}^\text{UL} = \SI{1.75}{eV}$ and $\zeta_{V,\text{incoh}}^\text{UL} = \SI{0.019}{eV}$ for $\phi \kpkm$ and $\zeta_{V,\text{coh}}^\text{UL} = \SI{0.47}{eV}$ and $\zeta_{V,\text{incoh}}^\text{UL} = \SI{0.025}{eV}$ for $\phi \ksks$ at the $90\%$ confidence level.

\section{Summary}

The processes $\epemtophikpkm$ and $\epemtophiksks$ have been studied for the first time using $22.7~\textrm{fb}^{-1}$ of electron-positron annihilation data taken at 33 different center-of-mass energies between 3.7730 GeV and 4.7008 GeV. The decay of the $\phi$ meson is clearly identified in both processes for all center-of-mass energies, and Born cross sections are determined with both high precision and accuracy.
No evidence for a resonant contribution is found from a fit to the $\epemtophikpkm$ and $\epemtophiksks$ Born cross sections. The upper limits at the $90\%$ confidence level for a wide range of resonance parameters $m$ and $\Gamma$ are set. This indicates that the $\psi(4230)$ strongly prefers to preserve its charm content in decays.

Since the continuum contributions to the Born cross sections for both processes $\epemtophikpkm$ and $\epemtophiksks$ are similar in shape, a constant fit is performed to its ratio yielding a proportionality factor of $3.85\pm0.01$. This result differs significantly from the value of two, thereby,  revealing an isospin symmetry breaking effect. Since the continuum production of the final states investigated goes through a virtual photon, isospin is a priori not conserved.


\begin{acknowledgments}
The BESIII collaboration thanks the staff of BEPCII and the IHEP computing center for their strong support. This work is supported in part by National Key R\&D Program of China under Contracts Nos. 2020YFA0406300, 2020YFA0406400; National Natural Science Foundation of China (NSFC) under Contracts Nos. 11635010, 11735014, 11835012, 11935015, 11935016, 11935018, 11961141012, 12022510, 12025502, 12035009, 12035013, 12061131003, 12192260, 12192261, 12192262, 12192263, 12192264, 12192265; the Chinese Academy of Sciences (CAS) Large-Scale Scientific Facility Program; the CAS Center for Excellence in Particle Physics (CCEPP); Joint Large-Scale Scientific Facility Funds of the NSFC and CAS under Contract No. U1832207; CAS Key Research Program of Frontier Sciences under Contracts Nos. QYZDJ-SSW-SLH003, QYZDJ-SSW-SLH040; 100 Talents Program of CAS; The Institute of Nuclear and Particle Physics (INPAC) and Shanghai Key Laboratory for Particle Physics and Cosmology; ERC under Contract No. 758462; European Union's Horizon 2020 research and innovation programme under Marie Sklodowska-Curie grant agreement under Contract No. 894790; German Research Foundation DFG under Contracts Nos. 443159800, 455635585, Collaborative Research Center CRC 1044, FOR5327, GRK 2149; Istituto Nazionale di Fisica Nucleare, Italy; Ministry of Development of Turkey under Contract No. DPT2006K-120470; National Research Foundation of Korea under Contract No. NRF-2022R1A2C1092335; National Science and Technology fund; National Science Research and Innovation Fund (NSRF) via the Program Management Unit for Human Resources \& Institutional Development, Research and Innovation under Contract No. B16F640076; Polish National Science Centre under Contract No. 2019/35/O/ST2/02907; Suranaree University of Technology (SUT), Thailand Science Research and Innovation (TSRI), and National Science Research and Innovation Fund (NSRF) under Contract No. 160355; The Royal Society, UK under Contract No. DH160214; The Swedish Research Council; U. S. Department of Energy under Contract No. DE-FG02-05ER41374
\end{acknowledgments}

%

\end{document}